# Microscopic Insights into Fatigue Mechanism in Wurtzite Ferroelectric $Al_{0.65}Sc_{0.35}N$: Oxygen Infiltration Enabled Grain Amorphization Spanning Boundary to Bulk

Ruiqing Wang[1], Danyang Yao[1, *], Jiuren Zhou[1, 2, *], Yang Li[1], Zhi Jiang[1], Dongliang Chen[1], Xu Ran[1], Yu Gao[1], Zixuan Cheng[1], Yong Wang[1], Yan Liu[1, 2], Yue Hao[1], Genquan Han[1, 2, *]

[1]School of Microelectronics, Xidian University, Xi'an, 710071, China; [2] Hangzhou Institute of Technology, Xidian University, Hangzhou, 311200, China. Email: {dyyao; zhoujiuren; gqhan}@xidian.edu.cn

Confronting the fatigue issue in wurtzite ferroelectrics, this work pioneers a study into the microscopic fatigue mechanism of ferroelectric AlScN film. Leveraging transmission electron microscopy and energy-dispersive spectrometry, we dynamically scrutinize the composition and distribution of nitrogen and oxygen elements in AlScN films during stress cycling. Prolonged cycling induces oxygen infiltration, initiating interactions at grain boundaries and permeating the bulk. This leads to series structural degradation, including nitrogen loss and ferroelectric grain amorphization, resulting in hard damage and ferroelectricity failure. The profound insights into the microscopic fatigue mechanism offer valuable guidance for the development of next-generation ferroelectric memories with superior endurance.

## I. Introduction

The increasing demand for high-capacity data storage in the digital era drives the exploration of next-generation non-volatile memory solutions with significant data volume and density [1]. Wurtzite ferroelectric memories, distinguished by a significant memory window, multi-level storage capability, and compatibility with back-end-of-line (BEOL) processes, stand out [2]. However, a critical endurance bottleneck, restricting cycles to approximately $10^3$ to $10^5$, poses a hurdle to its widespread adoption [3-7]. Despite continuous efforts to unravel the fatigue mechanism and enhance endurance, existing studies reveal persistent insufficiencies, necessitating a profound exploration of the microscopic fatigue mechanism.

In this study, we experimentally delve into the microscopic fatigue mechanism of $Al_{0.65}Sc_{0.35}N$ film, utilizing focused ion beam (FIB)-based transmission electron microscopy (TEM) and energy-dispersive spectrometry (EDS) techniques. The investigation unveils oxygen infiltration as the driving force behind grain amorphization and ferroelectricity failure, seamlessly spanning from the boundary to the bulk. This provides a clear understanding of the operational behavior and mechanism of oxygen infiltration during the fatigue process.

## II. Device Fabrication

Fig. 1(a) outlines the key process flow for fabricating AlScN-based capacitors. Commencing with a 150 nm thick Pt bottom electrode (BE) deposition on an AlN relaxed buffer layer, the subsequent step involves sputtering an AlScN film. This film, with an approximate 35% Sc concentration, is deposited at 400 °C, utilizing a $N_2$: Ar gas flow ratio of 9: 1. The top electrode (TE) consists of a 5 nm thick Ti layer and a 95 nm thick Au layer. The final step involves shadow mask the top electrode into a 200×200 μm² square shape.

## III. Results and discussion

TEM images in Fig. 1(b) reveal the complex structure of the fabricated device, featuring an AlScN layer with a thickness of 450 nm positioned between a Pt (BE) and a Ti/Au (TE). Fig. 1(c) plots its typical polarization (P) versus voltage (V) loops, evident confirming its ferroelectricity [8]. Further characterization using X-ray diffraction (XRD) and X-ray photoelectron spectroscopy (XPS) techniques (Figs. 2 and 3) enhances the understanding of AlScN film. The 2θ-θ curve scan on the (0002) reflection at around 36° and ω rocking scan of the AlScN (0002) with a FWHM value of 2.28° emphasize crystalline quality [9], while Sc concentration assessment reveals an approximate value of 35%.

In scrutinizing the fatigue characteristics and microscopic mechanisms of AlScN, our initial efforts to assess its endurance characteristics unveil remarkable performance, exceeding $2\times10^7$ cycles even under substantial stress (3.3 MV/cm amplitude, 50 μs period), surpassing reported counterparts [1-7] [Figs. 4-5]. This exceptional performance establishes a highly credible starting point for an in-depth analysis of its microscopic mechanisms.

Shifting our focus to the microscopic mechanism, TEM and EDS analysis were conducted on three samples: 1) Centre region of pristine device (**S1**); 2) Centre region of fatigue device after $1\times10^7$ cycle stress (**S2**); 3) Edge region of fatigue device after $1\times10^7$ cycle stress (**S3**). The comparison between S1 and S2 in Figs. 6(a) and (b) reveals that stress cycling drives oxygen to interact with AlScN film, infiltrating the bulk along the grain boundaries. In contrast to the electrode-covered centre region in S1 and S2, the partially electrode-covered edge region in S3, with an unlimited oxygen source, exhibits the phenomenon that stress cycling drives oxygen from the surface deep into the bulk, leading to nitrogen loss as well as oxidation and amorphization of the AlScN film.

Fig.7 quantitatively characterizes the oxygen content in all three samples. For the pristine device (S1), the average and localized oxygen content across the entire device is 2.3~2.5% [Fig. 7(a)]. After $10^7$ cycles of stress, S2 exhibits a substantial rise in oxygen content at grain boundaries, exceeding 10%, while internal grain regions maintain stability around 2.3%. These findings substantiate oxidation initiated at grain boundaries, likely triggered by the piezoelectric behavior of AlScN, creating a potential channel for oxygen infiltration [10]. Furthermore, Fig. 7(c) showcases a distinct crystalline state in S3, with the upper part fully oxidized and amorphous, resulting in an oxygen content exceeding 60%. Intriguingly, the bottom part maintains a well-preserved crystalline structure, with a slight decrease in oxygen content around 1.8%. Fig. 7(d) and (e) extracts the oxygen content of S1, S2, and S3, conclusively demonstrating that stress cycling propels oxygen infiltration from boundaries toward the entire bulk, ultimately causing film hard damage and ferroelectricity failure.

## IV. Conclusion

In a groundbreaking revelation, our study exposes the microscopic fatigue mechanism of ferroelectric AlScN. TEM and EDS techniques provide tangible evidence, vividly illustrating cycling-induced oxygen infiltration starting at boundaries and progressing into the bulk. This discovery sheds light on the endurance bottleneck in AlScN memories, providing crucial insights for future advancements.

**Acknowledgements:** This work was supported by the National Natural Science Foundation of China (Grant No. 62025402, 92264101, and 62090033), Major Program of Zhejiang Natural Science Foundation (Grant No. LDT23F0402).

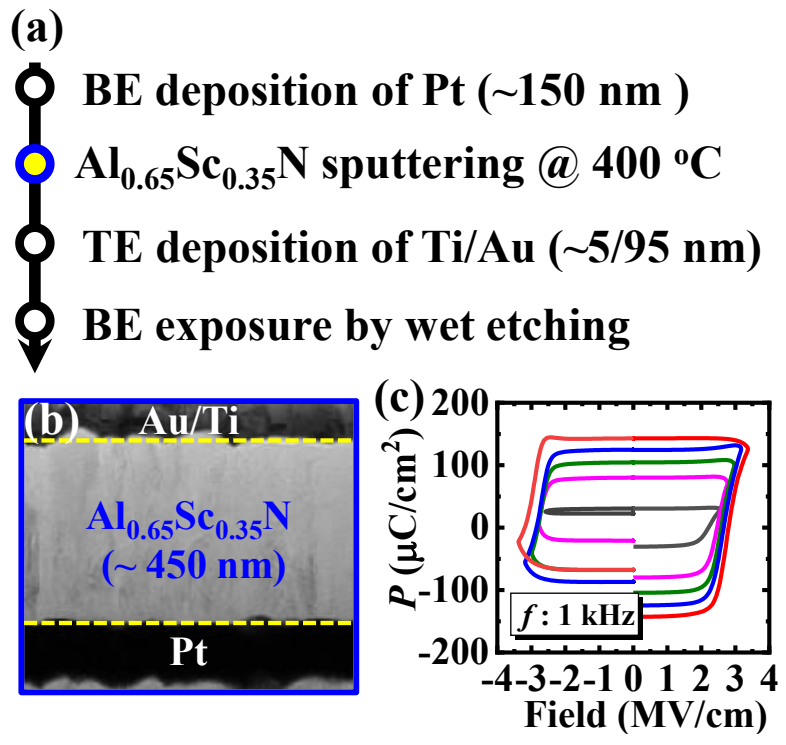

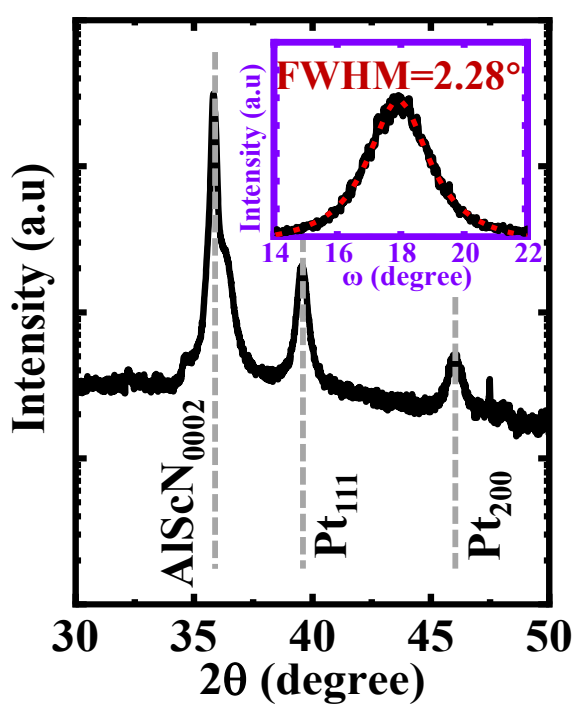

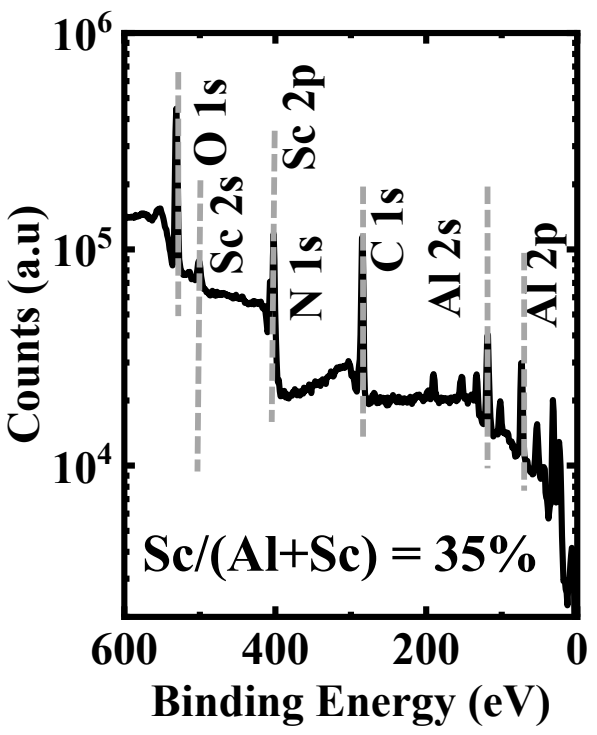

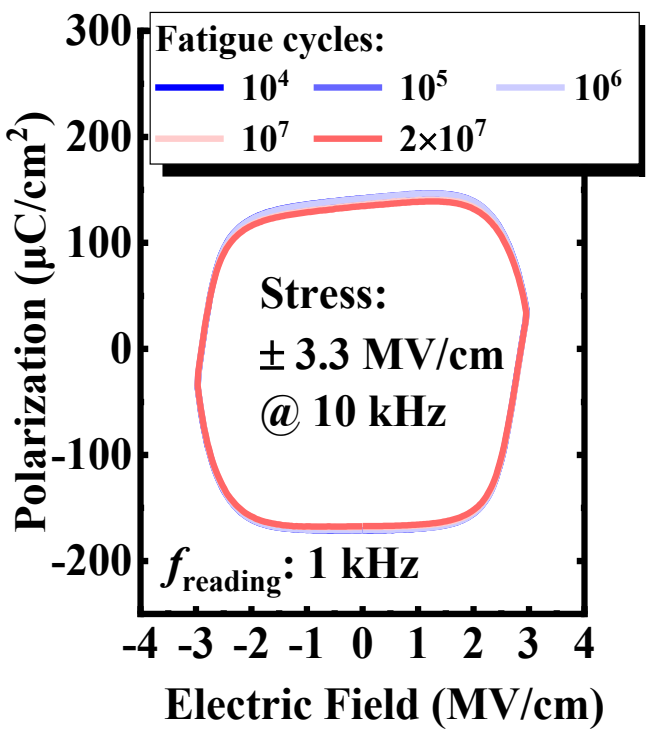

**Fig. 1.** (a) The key process flow for fabricating AlScN-based capacitors. (b) TEM images. (c) *P-V* loops.

**Fig. 2.** XRD results of fabricated AlScN with (0002) reflection, confirming its crystalline quality.

**Fig. 3.** XPS analysis of fabricated AlScN, delineating the Sc concentration of 35%.

**Fig. 4.** *P-E* loops concerning different number of stress cycles employing a wave of ±3.3 MV/cm @ 10 kHz.

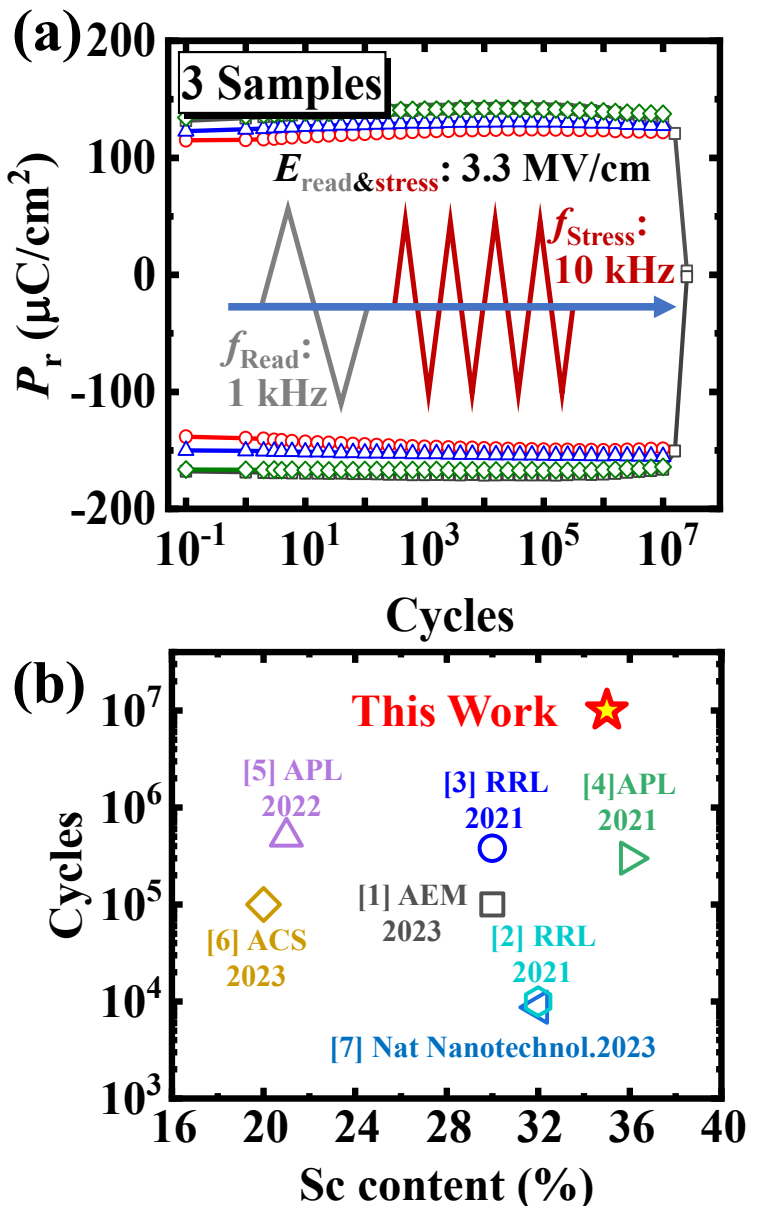

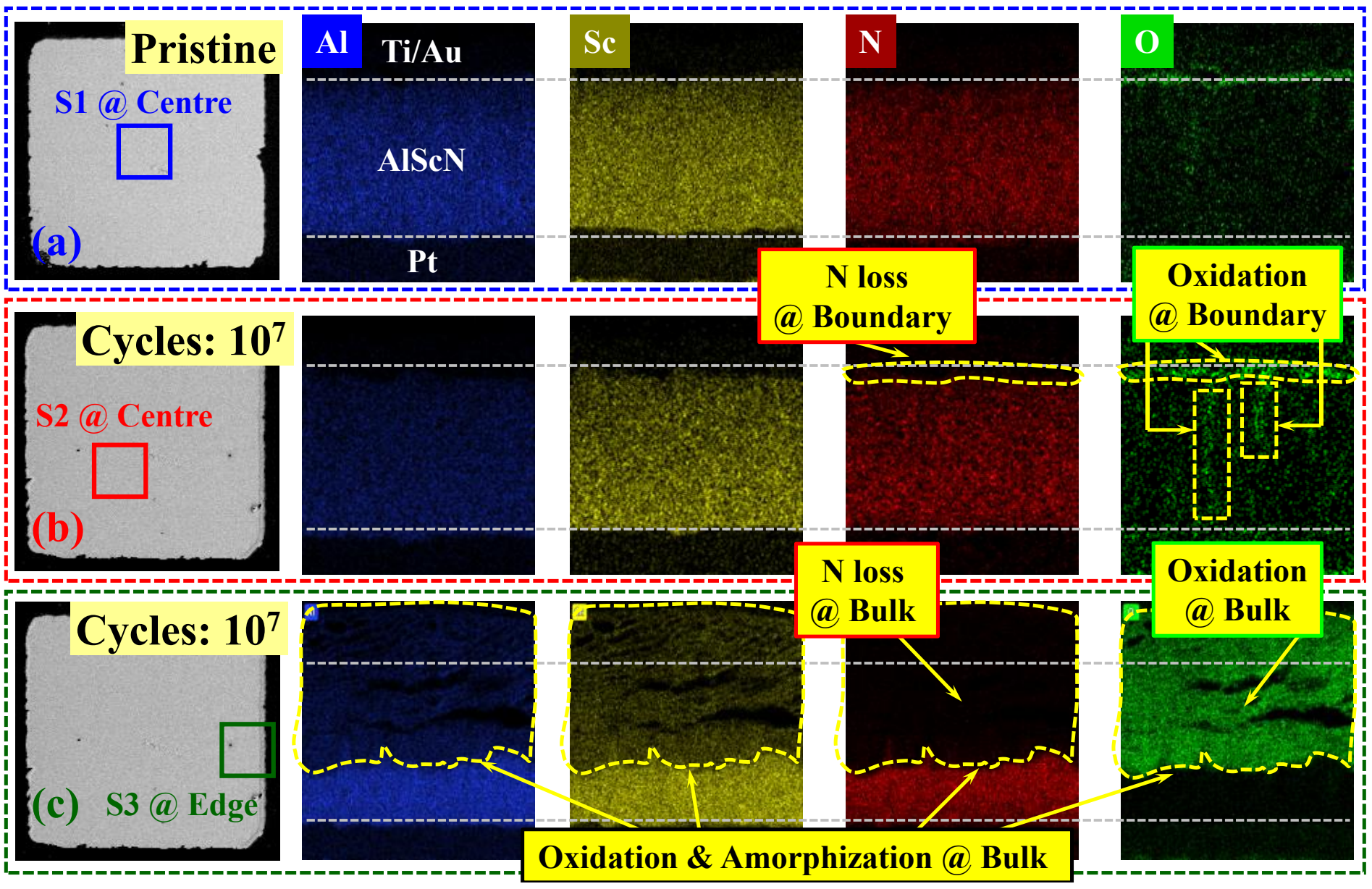

**Fig. 5.** Endurance characteristics of our AlScN film, exceeding $2\times10^7$ cycles, superior to all the reported counterparts.

**Fig. 6.** TEM and EDS analysis of three samples: 1) Centre region of pristine device (**S1**); 2) Centre region of fatigue device after $1\times10^7$ cycle stress (**S2**); Edge region of fatigue device after $1\times10^7$ cycle stress (**S3**), demonstrating stress cycling drives oxygen to interact with AlScN film, infiltrating the bulk along the grain boundaries, and further leading to nitrogen loss, oxidation, and amorphization of the AlScN film.

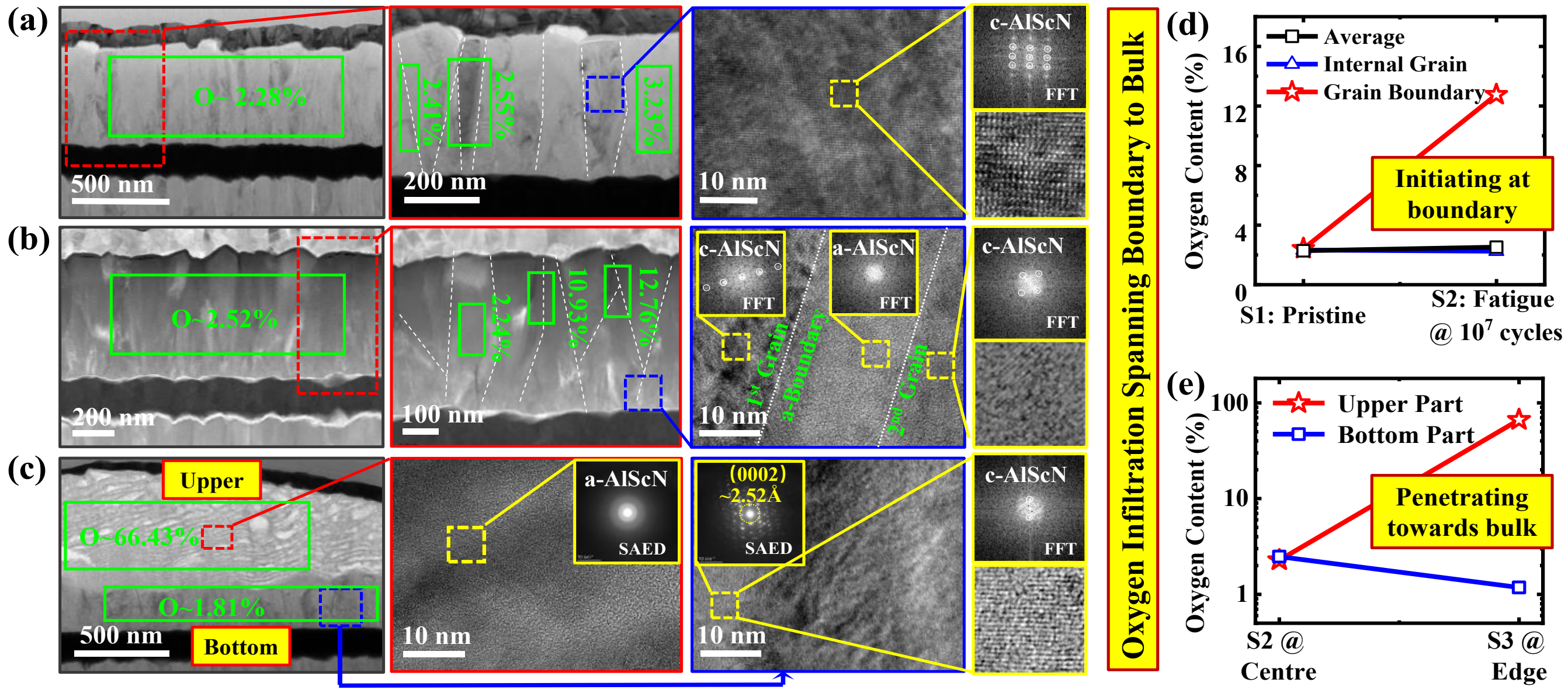

**Fig. 7.** Quantitative characterization of the oxygen content in all three samples, evidently confirming: 1) oxidation behavior initiated at grain boundaries, likely stemming from strain mismatch induced by piezoelectric behavior in AlScN, providing a potential pathway for oxygen infiltration; 2) stress cycling propels oxygen infiltration from boundaries toward the entire bulk, ultimately causing film hard damage and ferroelectricity failure.